\documentclass[sigconf]{acmart}
\AtBeginDocument{%
  \providecommand\BibTeX{{%
    \normalfont B\kern-0.5em{\scshape i\kern-0.25em b}\kern-0.8em\TeX}}}

\copyrightyear{2023}
\acmYear{2023}
\setcopyright{acmlicensed}\acmConference[RecSys '23]{Seventeenth ACM Conference on Recommender Systems}{September 18--22, 2023}{Singapore, Singapore}
\acmBooktitle{Seventeenth ACM Conference on Recommender Systems (RecSys '23), September 18--22, 2023, Singapore, Singapore}
\acmPrice{15.00}
\acmDOI{10.1145/3604915.3608857}
\acmISBN{979-8-4007-0241-9/23/09}

\usepackage{url}
\usepackage{color}
\usepackage{makecell}
\usepackage{subfigure}
\usepackage{makecell}
\usepackage{colortbl}
\usepackage{caption}
\usepackage{multirow}
\usepackage{enumitem}

\usepackage{wrapfig}
\usepackage{CJKutf8}

\newsavebox\wrbox
\AtBeginDocument{%
  \providecommand\BibTeX{{%
    \normalfont B\kern-0.5em{\scshape i\kern-0.25em b}\kern-0.8em\TeX}}}

\newcommand{\ie}{\emph{i.e., }}
\newcommand{\eg}{\emph{e.g., }}

\begin{document}


\title{TALLRec: An Effective and Efficient Tuning Framework to Align Large Language Model with Recommendation}


\author{Keqin Bao*}
\email{baokq@mail.ustc.edu.cn}
\affiliation{%
  \institution{University of Science and Technology of China}
  \country{China}
}

\author{Jizhi Zhang*}
\email{cdzhangjizhi@mail.ustc.edu.cn}
\affiliation{
  \institution{University of Science and Technology of China}
  \country{China}
}

\author{Yang Zhang}{
\email{zy2015@mail.ustc.edu.cn}
\affiliation{
  \institution{University of Science and Technology of China}
    \country{China}
}
}

\author{Wenjie Wang}{
\email{wenjiewang96@gmail.com}
\affiliation{
  \institution{National University of Singapore}
    \country{Singapore}
}
}

\author{Fuli Feng\dag}{
\email{fulifeng93@gmail.com}
\affiliation{
  \institution{University of Science and Technology of China}
    \country{China}
}
}

\author{Xiangnan He\dag}{
\email{xiangnanhe@gmail.com}
\affiliation{
  \institution{University of Science and Technology of China}
    \country{China}
}
}

\thanks{*The two authors contributed equally to this work and are listed alphabetically.~\dag The corresponding author.}




\begin{abstract}
Large Language Models (LLMs) have demonstrated remarkable performance across diverse domains, thereby prompting researchers to explore their potential for use in recommendation systems. Initial attempts have leveraged the exceptional capabilities of LLMs, such as rich knowledge and strong generalization through In-context Learning, which involves phrasing the recommendation task as prompts. Nevertheless, the performance of LLMs in recommendation tasks remains suboptimal due to a substantial disparity between the training tasks for LLMs and recommendation tasks, as well as inadequate recommendation data during pre-training. To bridge the gap, we consider building a Large Recommendation Language Model by tunning LLMs with recommendation data. To this end, we propose an efficient and effective Tuning framework for Aligning LLMs with Recommendations, namely TALLRec. We have demonstrated that the proposed TALLRec framework can significantly enhance the recommendation capabilities of LLMs in the movie and book domains, even with a limited dataset of fewer than 100 samples. Additionally, the proposed framework is highly efficient and can be executed on a single RTX 3090 with LLaMA-7B. Furthermore, the fine-tuned LLM exhibits robust cross-domain generalization. Our code and data are available at \url{https://github.com/SAI990323/TALLRec}.
\end{abstract}





\begin{abstract}
Large Language Models (LLMs) have demonstrated remarkable performance across diverse domains, thereby prompting researchers to explore their potential for use in recommendation systems. Initial attempts have leveraged the exceptional capabilities of LLMs, such as rich knowledge and strong generalization through In-context Learning, which involves phrasing the recommendation task as prompts. Nevertheless, the performance of LLMs in recommendation tasks remains suboptimal due to a substantial disparity between the training tasks for LLMs and recommendation tasks, as well as inadequate recommendation data during pre-training. To bridge the gap, we consider building a Large Recommendation Language Model by tunning LLMs with recommendation data. To this end, we propose an efficient and effective Tuning framework for Aligning LLMs with Recommendations, namely TALLRec. We have demonstrated that the proposed TALLRec framework can significantly enhance the recommendation capabilities of LLMs in the movie and book domains, even with a limited dataset of fewer than 100 samples. Additionally, the proposed framework is highly efficient and can be executed on a single RTX 3090 with LLaMA-7B. Furthermore, the fine-tuned LLM exhibits robust cross-domain generalization. Our code and data are available at \url{https://github.com/SAI990323/TALLRec}.
\end{abstract}

\maketitle
\section{Introduction}
\label{introduction}
\textit{Large Language Models} (LLMs) have exhibited remarkable proficiency in generating text that closely resembles human language and in performing a wide range of tasks~\cite{LLM_survey}, including Natural Language Processing~\cite{liang2022holistic, chowdhery2022palm, wei2022emergent}, Robotics~\cite{LLM_rob_1, LLM_rob_2, LLM_rob_3}, and Information Retrieval~\cite{LLM_IR_1, LLM_IR_2, LLM_IR_3, ai2023information}.
Prior research has also demonstrated the knowledge-rich and compositional generalization capabilities of LLMs~\cite{ouyang2022training,sanh2021multitask,wei2021finetuned}. 
Only given appropriate instructions, these models are able to learn how to solve unseen tasks and inspire their own knowledge to achieve a high level of performance~\cite{lin2021few}.
The aforementioned capabilities of LLM present promising opportunities to address the current challenges requiring strong generalization and rich knowledge in the recommendation field.
In this light, it is valuable to explore the integration of LLMs into recommender systems, which has received limited attention in prior research.

\begin{figure*}[t]
  \centering
  \includegraphics[width=\linewidth]{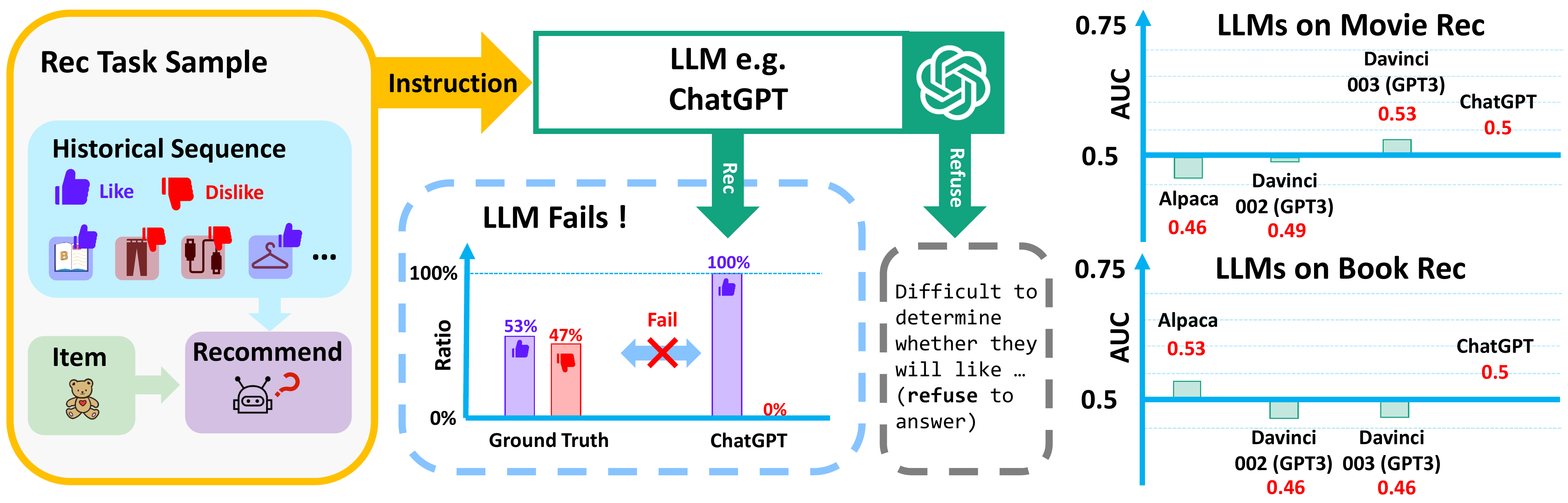}
  \caption{Illustration of LLMs for the recommendations. Given users' interaction history, LLMs predict whether a user will like a new item through In-context Learning. 
  However, the representative LLMs, \eg ChatGPT, either refuse to answer or always give positive predictions (likes) on Movie and Book recommendation tasks. 
  If we ignore the refused answers and calculate AUC on the remaining samples, we find that LLMs perform similarly with random guessing (AUC=0.5). Refer to Section~\ref{sec:exp} for more experimental details. 
  }

  \label{fig:llm_fail}
\end{figure*}

In recent initial attempts~\cite{gao2023chat, wang2023zero}, achieving the target relies on \textit{In-context Learning}~\cite{brown2020language}, which is typically implemented through the official OpenAI API~\cite{openaiapi}.
They regard the LLM as a toolformer~\cite{schick2023toolformer} of traditional recommendation models (such as MF~\cite{koren2009matrix} and LightGCN~\cite{he2020lightgcn}), \ie the LLM is used for re-ranking the candidate items filtered by these models.
However, these approaches only reach a comparable performance with traditional models~\cite{gao2023chat, wang2023zero}.
Worse still, using only In-context Learning may fail to make recommendations. As shown in Figure~\ref{fig:llm_fail}, we find that ChatGPT either refuses to answer or always gives positive predictions (likes). 
Therefore, it is critical to further explore an appropriate way for more effective leverage of LLMs in the recommendation.

We postulate that the failure of using only In-context Learning is because of two reasons: 1) LLMs may not align well with the recommendation task due to the huge gap between language processing tasks for training LLMs and recommendation. Besides, the recommendation-oriented corpus is very limited during the training phase of LLMs.
2) The effect of LLMs is restricted by the underlying recommendation models, which may fail to include target items in their candidate lists due to their limited capacity.
Therefore, we consider building a \textit{Large Recommendation Language Model} (LRLM) to bridge the gap between LLMs and the recommendation task and better stimulate the recommendation capabilities of LLMs in addition to In-context Learning.

Toward this goal, we focus on tuning LLMs with the recommendation task.
Considering that instruction tuning is core to letting the LLM learn to solve different tasks and have strong generalization ability~\cite{instruct_gen_1, instruct_gen_2, instruct_gen_3}, we propose a lightweight tuning framework to adapt LLMs for recommendations, named TALLRec. 
Elaborately, TALLRec structures the recommendation data as instructions and tunes the LLM via an additional instruction tuning process. 
Moreover, given that LLM training necessitates a substantial amount of 
computing resources, 
TALLRec employs a lightweight tuning approach to efficiently adapt the LLMs to the recommendation task.

Specifically, we apply the TALLRec framework on the LLaMA-7B model~\cite{llama} with a LoRA~\cite{lora} architecture, which ensures the framework can be deployed on an Nvidia RTX 3090 (24GB) GPU. 
Furthermore, to investigate the minimal computational resources required, we do experiments in a few-shot setting, utilizing only a limited number of tuning examples.
We conduct detailed experiments in knowledge-rich recommendation scenarios of movies and books, where the tuned LLaMA-7B model outperforms traditional recommendation models and In-context Learning with GPT3.5, a much stronger LLM than LLaMA-7B.
The results validate the efficiency and robustness of our framework: 1) our TALLRec framework can quickly inspire the recommendation capability of LLMs in the few-shot setting.
and 2) LLMs trained via the TALLRec framework have a strong generalization ability 
across different domains (\eg \textit{movie} $\rightarrow$ \textit{book}).

In total, our contributions are summarized as follows:
\begin{itemize}

    \item We study a new problem in recommendation --- aligning the LLMs with the recommendation, where we reveal the limitations of In-context Learning-based approaches and underscore the significance of instruction tuning.
    \item We introduce a new TALLRec framework to build Large Recommendation Language Models, which enables the effective and efficient tuning of LLMs for recommendation with low GPU costs and few tuning samples.
    \item We conduct extensive experiments, validating the effectiveness and efficiency of the proposed framework, and uncovering its exceptional robustness with seamless navigation across different domains.
\end{itemize}

\section{TALLRec}
In this section, we first introduce the preliminary knowledge for tuning LLMs and our task formulation, and then present the proposed TALLRec framework.

\subsection{Preliminary}
\label{section:preliminary}

\begin{table}
\vspace{-0.5cm}
\centering
 \tiny  
  \caption{ 
 A tuning sample for a translation task.
    }
 \label{table-operators}
 \vspace{-0.4cm}
 \resizebox{0.4\textwidth}{!}{
    \begin{tabular}{ll}
        \toprule
        \multicolumn{2}{c}{\textbf{Instruction Input}} \\
        \midrule
        {Task Instruction:}  & Translate from English to Chinese.\\ 
        {Task Input:} & Who am I ? \\  
        \midrule
        \multicolumn{2}{c}{\textbf{Instruction Output}}  \\
        \midrule
        {Task Output:}  &  \begin{CJK}{UTF8}{gbsn}我是谁?\end{CJK} \\ 
        \bottomrule
    \end{tabular}
}
\end{table}
\textbf{$\bullet$ {Instruction Tuning}} is a crucial technique to train LLMs with human-annotated instructions and responses~\cite{ouyang2022training}. 
Generally, instruction tuning has four steps (see the example in Table~\ref{table-operators}). Specifically, \textbf{Step 1:} Define a task and articulate a ``\textit{Task Instruction}'' using natural language, which usually encompasses a clear definition of the task, as well as specific solutions to address it. 
\textbf{Step 2:} Formulate and construct the input and output of the task in natural language, denoted as ``\textit{Task Input}'' and ``\textit{Task Output}''.
\textbf{Step 3:} Integrate the ``\textit{Task Instruction}'' and ``\textit{Task Input}'' together to form the ``\textit{Instruction Input}'', and take the ``\textit{Task Output}'' as the corresponding ``\textit{Instruction Output}'', for each tuning sample. 
\textbf{Step 4:} Instruction tuning on LLMs based on the formatted pairs of ``\textit{Instruction Input}'' and ``\textit{Instruction Output}''.

\begin{table}
\centering
 \tiny  
  \caption{A tuning sample for rec-tuning.}
 \label{table-operators}
 \resizebox{0.45\textwidth}{!}{
    \begin{tabular}{@{}ll@{}}
        \toprule
        \multicolumn{2}{c}{\textbf{Instruction Input}} \\
        \midrule
        {Task Instruction:}  & \makecell[lp{4cm}]{Given the user's historical interactions, please determine  whether the user will enjoy the target new movie by answering "Yes" or "No".}\\ 
        \midrule
        {Task Input:} & \makecell[lp{4cm}]{User's liked items: GodFather. \\ User's disliked items: Star Wars.\\Target new  movie:  Iron Man}\\
        \midrule
        \multicolumn{2}{c}{\textbf{Instruction Output}}  \\
        \midrule
        
        {Task Output:}  &  No. \\ 
        \bottomrule
  \label{table:rec-tuning}
    \end{tabular}
    }
\end{table}

\vspace{5pt}
\noindent\textbf{\textbf{$\bullet$ Rec-tuning Task Formulation.}}
We aim to utilize LLM, denoted as $\mathcal{M}$, to construct an LRLM, which can predict whether a new item will be enjoyed by a user. 
To achieve this objective, we do recommendation tuning (rec-tuning) on LLMs with recommendation data.
As shown in Table~\ref{table:rec-tuning}, we format recommendation data into a pattern of instruction tuning. 
We begin by composing a ``\textit{Task Instruction}'' that directs the model to determine whether the user will like the target item based on their historical interactions, and to respond with a binary answer of ``Yes'' or ``No''. 
To format the ``\textit{Task Input}'', we categorize the user's historically interacted items into two groups based on ratings: 
the user's liked items and disliked items, where items are sequentially ranked by interaction time and represented by textual descriptions (\eg title and brief introduction). 
Besides, ``\textit{Task Input}'' also includes a target new item that the user has never seen. 
Lastly, we merge ``\textit{Task Instruction}'' and ``\textit{Task Input}'' to create a ``\textit{Instruction Input}'', and then set the expected ``\textit{Instruction Output}'' as `Yes'' or ``No'' for rec-tuning.

\subsection{TALLRec Framework}

\begin{figure*}[t]
  \centering
  \includegraphics[width=\linewidth]{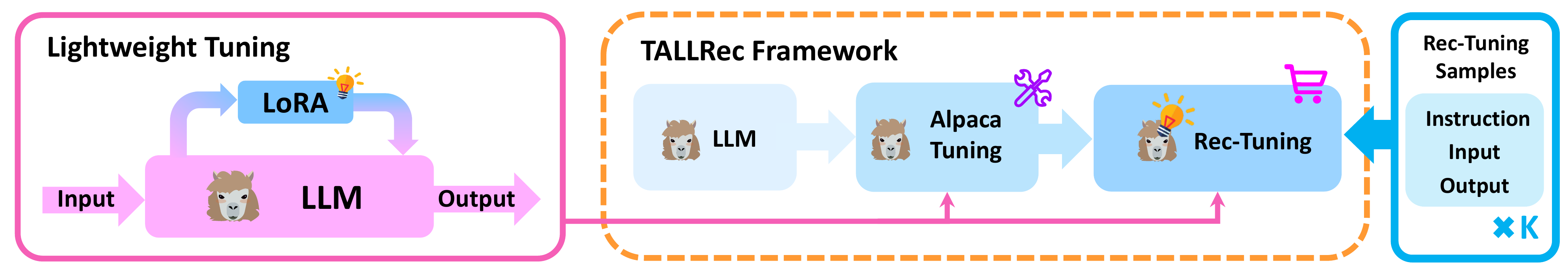}
  \caption{Illustration of the TALLRec framework constructed by alpaca tuning and rec-tuning two stages. During rec-tuning, we use the rec-tuning samples with instruction input and output constructed from recommendation data. Notably, we employ lightweight tuning technology to enhance the efficiency of our TALLRec framework.}
  \label{fig:frame_work}
\end{figure*}

In this subsection, we introduce the TALLRec framework, which aims to facilitate the effective and efficient alignment of LLMs with recommendation tasks, particularly in low GPU memory consumption settings. 
Specifically, we first present two TALLRec tuning stages with lightweight implementation, followed by the backbone selection. As shown in Figure~\ref{fig:frame_work}, TALLRec comprises two tuning stages: alpaca tuning and rec-tuning. The former stage is the common training process of LLM that enhances LLM's generalization ability, while the latter stage emulates the pattern of instruction tuning and tunes LLMs for the recommendation task.

\vspace{+5pt}
\noindent\textbf{$\bullet$ TALLRec Tuning Stages.}
For alpaca tuning, we employ the self-instruct data made available by Alpaca~\cite{alpaca} to train the LLM. 
Specifically, we utilize the conditional language modeling objective during the alpaca tuning, as exemplified in the Alpaca repository\footnote{\url{https://github.com/tloen/alpaca-lora}.}. Formally,
\begin{equation}\small
    \max_{\Phi} \sum_{(x,y)\in\mathcal{Z}} \sum_{t=1}^{|y|}  \text{log} \left(  P_{\Phi}(y_{t} | x, y_{<t}) \right),
\end{equation}
where $x$ and $y$ represent the ``\textit{Instruction Input}'' and ``\textit{Instruction Output}'' in the self-instruct data, respectively, $y_t$ is the $t$-th token of the $y$, $y_{<t}$ represents the tokens before $y_{t}$,
$\Phi$ is the original parameters of $\mathcal{M}$, and $\mathcal{Z}$ is the training set. For rec-tuning, we can leverage the rec-tuning sample as described in Table~\ref{table:rec-tuning} to tune the LLM, similar to alpaca tuning.

\vspace{5pt}
\noindent{\textbf{$\bullet$ Lightweight Tuning.}} 
However, directly tuning the LLM is computationally intensive and time-consuming.
As such, we propose to adopt a lightweight tuning strategy to execute both alpaca tuning and rec-tuning. 
The central premise of lightweight tuning is that contemporary language models may possess an excessive number of parameters, and their information is concentrated on a low intrinsic dimension~\cite{lora}. Consequently, we can achieve comparable performance to that of the entire model by tuning only a small subset of parameters~\cite{prefix, adapter, prompt_tuning}.
Specifically, we employ LoRA~\cite{lora}, which involves freezing the pre-trained model parameters and introducing trainable rank decomposition matrices into each layer of the Transformer architecture to facilitate lightweight tuning.
Therefore, by optimizing rank decomposition matrices, we can efficiently incorporate supplementary information while maintaining the original parameters in a frozen state.
In total, the final learning objective can be computed as:
\begin{equation}\small
    \max_{\Theta} \sum_{(x,y)\in\mathcal{Z}} \sum_{t=1}^{|y|}  \text{log} \left(  P_{\Phi + \Theta}(y_{t} | x, y_{<t}) \right),
\end{equation}
where $\Theta$ is the LoRA parameters and we only update LoRA parameters during the training process. 
Through LoRA, we can complete training with only one-thousandth of the original LLM parameters to complete the training process~\cite{lora}.
\vspace{5pt}
\noindent{\textbf{$\bullet$ Backbone Selection}.}
At present, there are large amounts of LLMs released, such as GPT series, PaLM, CHinchilla, and LLaMA~\cite{brown2020language, chowdhery2022palm, chinchilla, llama}.
Among these, a considerable number of LLMs (such as PaLM and Chinchilla) do not provide access to their model parameters or APIs, rendering them challenging to utilize for research or other applications. 
Additionally, data security concerns are significant issues in the recommendation field.
Consequently, the utilization of third-party APIs (such as ChatGPT and text-davinci-003) to leverage LLMs necessitates further discussion.
To replicate the issues that require consideration in real-world recommendation scenarios, we intend to simulate the practical utilization of a public LLM and update its parameters for recommendation purposes.
After careful consideration, we have opted to conduct experiments using LLMs-LLaMA, which is presently the best-performing open-source LLM, and whose training data is also publicly available~\cite{llama}.
\section{Experiments}\label{sec:exp}
In this section, we conduct experiments to answer the following research questions:
\begin{itemize}[leftmargin=*]
    \item [-] \textbf{RQ1:} 
    How does TALLRec perform compared with current LLM-based and traditional recommendation models? 
    \item [-] \textbf{RQ2:} 
    How do the different components in TALLRec affect its effectiveness?
    \item [-] \textbf{RQ3:} 
    How does TALLRec perform under cross-domain recommendation?
\end{itemize}

\begin{table*}[t]
\setlength{\abovecaptionskip}{0cm}
\setlength{\belowcaptionskip}{0cm}
\caption{
Performance comparison between conventional sequential recommendation baselines and TALLRec under different few-shot training settings. The reported result is the AUC multiplied by 100, with boldface indicating the highest score. $\ddagger$: significantly better than all baselines with t-test $p$<0.01.
}
\setlength{\tabcolsep}{4mm}{
\resizebox{\textwidth}{!}{
\begin{tabular}{ccccccccc}
\toprule
 & \colorbox{gray!10}{Few-shot} & GRU4Rec                  & Caser                & SASRec               & DROS  & GRU-BERT & DROS-BERT &        TALLRec \\
 \midrule
 \multirow{3}{*}{movie} & 16   & 49.07                & 49.68                & 50.43                & 50.76    &50.85 & 50.21 & \textbf{67.24}$\ddagger$\\
                        & 64   & 49.87                & 51.06                & 50.48                 & 51.54    &51.65  & 51.71 &  \textbf{67.48}$\ddagger$\\
                        & 256  & 52.89                & 54.20                & 52.25                 & 54.07    &53.44 & 53.94 & \textbf{71.98}$\ddagger$\\ 
\midrule
\multirow{3}{*}{book}   & 16     & 48.95              & 49.84                & 49.48                & 49.28   & 50.07 & 50.07 &  \textbf{56.36}\\
                        & 64     & 49.64              & 49.72                & 50.06                & 49.13  & 49.64 & 48.98 & \textbf{60.39}$\ddagger$ \\
                        & 256    & 49.86              & 49.57                & 50.20                & 49.13  & 49.79 & 50.20 & \textbf{64.38}$\ddagger$\\ \bottomrule
\end{tabular}
}}

\label{tab:over_all_conpare}
\end{table*}

\begin{figure*}[t]
\centering 
\includegraphics[width=\textwidth]{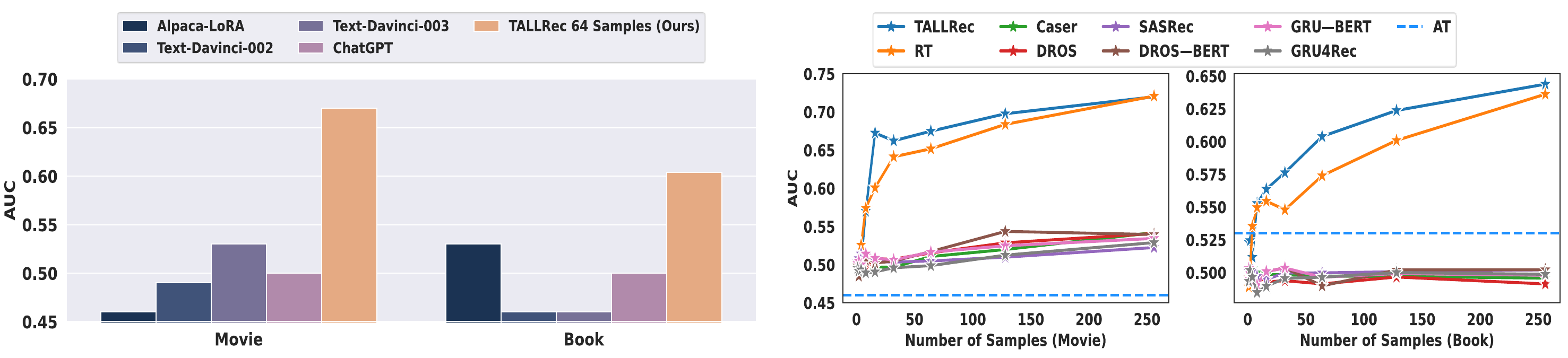}
\caption{
 Figure (a) shows the performance comparison between LLM-based baselines (zero-shot setting) and ours TALLRec, where the TALLRec is trained on only 64 rec-tuning samples  (\textit{i.e.}, in the 64-shot training setting). Figure (b) demonstrates the performance tendency of TALLRec's variants and conventional sequential recommendation methods \textit{w.r.t.} the number of training samples used, ranging from 1 to 256. TALLRec has three variants: ``AT'' for alpaca tuning only, ``RT'' for rec-tuning only, and ``TALLRec'' for the full version.
}
\label{fig:LLM_result}
\end{figure*} 

\noindent\textbf{$\bullet$ Dataset.}
We conduct experiments on two datasets. The statistics and more details can be found in our released data. 
\begin{itemize}[leftmargin=*]
    \item [-] \textbf{Movie.} 

    This is a processed dataset from MovieLens100K~\cite{ml-100k}
    , which comprises user ratings on movies and comprehensive textual descriptions of movies such as ``title'' and ``director''. Because we conduct experiments in a few-shot training setting that requires limited tuning samples, 
    we process the original dataset by sampling the most recent 10,000 interactions and split them into training, validation, and testing sets with a ratio of 8:1:1. 
    To construct a rec-tuning sample, 10 interactions prior to the target item are retained as historical interactions. 
    Following~\cite{he2020lightgcn, zhang2023reformulating}, we only treat the interactions with ratings $>3$ as ``likes'', and those with ratings $\leq 3$ as ``dislike''. 

    \item [-] \textbf{Book.} 
    This is a book recommendation dataset processed from BookCrossing~\cite{book_crossing}. The BookCrossing dataset has user ratings (1-10) and textual descriptions of books, such as the information of `Book-Author' and `Book-Title'. For each user, we randomly select an item interacted by this user as the prediction target, and sample $10$ interacted items as historical interactions\footnote{BookCrossing lacks interaction timestamps, thus we can only construct historical interaction by random sampling.}. Subsequently, we partition constructed rec-tuning samples into training, validation, and testing sets with the same ratio of 8:1:1. Additionally, we binarize the ratings according to a threshold of 5. 

\end{itemize}


\noindent\textbf{$\bullet$ Few-shot Training Setting.}
We adopt a few-shot training setting, where only a limited number of samples are randomly selected from the training set for model training. It is referred to as `$K$-shot' training setting, where $K$ represents the number of training samples used. By setting an extremely small value for K, such as 64, we could test whether a method can rapidly acquire recommendation capability from LLMs with severely limited training data.

\vspace{+5pt}
\noindent\textbf{$\bullet$ Baseline.}
We compare TALLRec against both LLM-based and traditional recommendation methods. 1) Existing \textbf{LLM-based methods} adopt In-context Learning to directly generate recommendations~\cite{gao2023chat, wang2023zero}. For a fair comparison, we align these methods with TALLRec by using the same instructions. Specifically, we perform In-context Lerning on  different LLMs: 1) \textit{Alpaca-LoRA}, 2) \textit{Text-Davinvi-002}, 3) \textit{Text-Daviniv-003}, and 4) \textit{ChatGPT}. Alpaca-LoRA is a model for reproducing Alpaca results of the LLaMA model by using LoRA and alpaca tuning. The latter three are GPT series models from OpenAI.

2) \textbf{Traditional methods.} Since our approach utilizes historical interactions to predict the subsequent interaction, similar to the sequential recommendation, we consider the following sequential models: 
\textbf{(i) GRU4Rec~\cite{GRU4Rec}} is an RNN-based sequential recommender, which utilizes GRU to encode historical interactions.
\textbf{(ii) Caser~\cite{Caser}} utilizes CNN to encode historical interaction sequences. 
\textbf{(iii) SASRec \cite{SASRec}} is a classic transformer-based sequential recommender. 
\textbf{(iv) DROS \cite{DROS}} is a state-of-the-art sequential recommender model, which harnesses distributionally robust optimization for robust recommendations. We use the version implemented by GRU4Rec, provided by the authors.\footnote{\url{https://github.com/YangZhengyi98/DROS}.}
The sequential models above rely on item ID features without considering textual descriptions of items. However, in our setting, we assume item text descriptions are available for LLM tuning. To ensure fair comparisons, we further consider comparing the following variants of GRU4Rec and DROS: 
\textbf{(v) GRU-BERT} is a variant of GRU4Rec that incorporates a pre-trained BERT~\cite{devlin-etal-2019-bert} to encode text descriptions. Specifically, BERT encodes text descriptions and outputs a CLS embedding, which is then concatenated with the original ID embeddings of GRU4Rec as the item representations. 
\textbf{(vi) DROS-BERT} is integrated with BERT, similar to GRU-BERT. 

\vspace{+5pt}
\noindent\textbf{$\bullet$ Evaluation Metric.}
Since TALLRec aims to predict user preference over a given target item, \ie a binary classification problem, we adopt a popular evaluation metric used in recommendation: Area Under the Receiver Operating Characteristic (AUC).

\vspace{+5pt}
\noindent\textbf{$\bullet$ Implementation Details.}
To ensure uniform sequence lengths, we use the user's last interacted item to pad the historical interaction sequences with lengths $<$ the threshold, 10. 
For all methods, we optimize parameters using Adam with MSE loss and a learning rate of 1e-3. We search the weight decay of all methods in \{1e-3, 1e-4, 1e-5, 1e-6, 1e-7\}. 
Following~\cite{DROS}, regarding specific hyperparameters of baselines, we adhered to their original settings. 
For GRU-BERT and DROS-BERT, we utilize BERT released by Hugging Face\footnote{\url{https://huggingface.co/bert-base-uncased}.}, while setting the number of GRU layers to $4$ and the hidden size to 1024 for aligning with BERT's embedding size.
Lastly, we run all methods five times with different random seeds and report the averaged results.

\subsection{Performance Comparison (RQ1)}
We aim to investigate the recommendation performance of various methods under the few-shot training setting, which enables us to evaluate 
how LLMs can be quickly adjusted for recommendation with limited rec-tuning samples. 
The evaluation results against traditional methods are presented in Table~\ref{tab:over_all_conpare}, while the comparison against LLM-based methods is depicted in Figure~\ref{fig:LLM_result} (a).

From the figure and table, we draw the following observations: 
1) Our method significantly outperforms both traditional and LLM-based methods, verifying the superiority of tuning LLM via our TALLRec framework. TALLRec successfully unlocks the knowledge and generalization capabilities of LLMs for recommendations.
2) LLM-based methods perform similarly to random guessing (AUC$\approx$0.5). However, the LLMs trained via TALLRec achieves significant improvements. These results demonstrate a considerable gap between recommendation and language tasks, showing the importance of using recommendation data for rec-tuning on LLMs. 
3) Traditional recommender methods consistently yield inferior performance under our few-shot training settings, implying that traditional methods are incapable of quickly learning the recommendation capability with limited training samples.
4) GRU-BERT and DROS-BERT do not show significant improvement over their backend models, GRU4Rec and DROS. This indicates that purely adding textual descriptions cannot enhance the traditional recommender models in the few-shot training setting.

\begin{figure*}[t]
\centering  
\includegraphics[width=\textwidth]{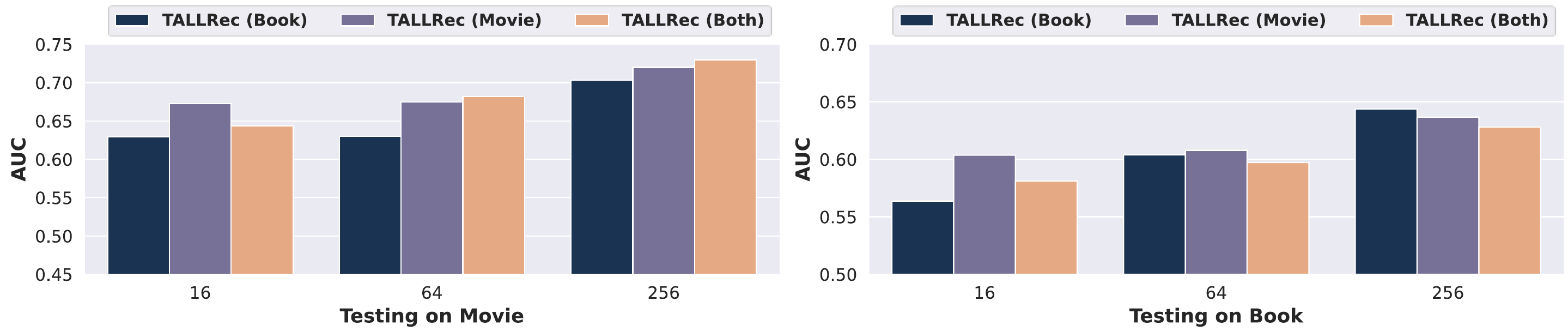}
\caption{
Cross-domain generalization performance of LRLMs trained via TALLRec using Book data (TALLRec (Book)), Movie data (TALLRec (Movie)), and both (TALLRec (Both)). The left figure shows the testing results on the Movie dataset with varying numbers of rec-tuning samples, while the right figure shows the testing results on the Book dataset.
}
\vspace{-5pt}
\label{fig:over_cross_domain}
\vspace{-5pt}
\end{figure*}

\subsection{Ablation Study (RQ2)}

To demonstrate the effectiveness of alpaca tuning and rec-tuning in TALLRec, we conduct ablation studies with varying $K$ under the $K$-shot training setting. 
Specifically, we compare the performance of TALLRec with that of two variants, ``AT'' and ``RT'', where ``AT'' only conducts the alpaca tuning, while ``RT'' solely implements rec-tuning. By varying $K$, we further investigate the impact of the number of training samples.

We summarize the results in Figure~\ref{fig:LLM_result} (b), from which we have the following observations:
1) The performance of ``AT'' significantly declines compared to that of ``RT'' and TALLRec, indicating the essential effect of rec-tuning, which effectively inspires the LLM's recommendation capability
2) With limited rec-tuning samples ($\leq128$), TALLRec generally outperforms ``RT'', confirming that alpaca tuning can enhance the LLM's generalization ability on new tasks, especially when the training data in the new tasks are insufficient. 
As the quantity of rec-tuning samples grows, the results of TALLRec and ``RT'' become closer. This makes sense, as the significance of generalization abilities derived from other tasks diminishes when there is an ample amount of training data for the new tasks. 
3) With the increase of rec-tuning sample number, TALLRec consistently performs better than the baselines. It is attributed to rec-tuning, which can utilize limited samples to inspire the LLM's recommendation capability. 

\subsection{Cross-domain Generalization Analyses (RQ3)}
To further investigate the generalization ability of TALLRec, we conduct experiments on cross-domain recommendations. 
Specifically, we tune TALLRec with different rec-tuning samples, including 1) ``TALLRec (Book)'', only using the samples from the Book dataset; 2) ``TALLRec (Movie)'', solely using samples from the Movie dataset; and 3) ``TALLRec (Both)'', tuned with both the Book and Movie samples. 
We vary $K$ in $\{16, 64, 258\}$ under the few-shot training setting, and evaluate the models on the testing sets of Book and Movie, respectively. 
The results are summarized in Figure~\ref{fig:over_cross_domain}, from which we can find: 
1) TALLRec demonstrates remarkable cross-domain generalization ability. For instance, after tuning only on movie samples, ``TALLRec (Movie)'' exhibits good performance on Book data, comparable to ``TALLRec (Book)''. This is impressive and suggests that TALLRec has cross-domain generalization ability instead of only fitting a single domain like traditional recommenders.
2) ``TALLRec (Both)'' surpasses ``TALLRec (Movie)'' and ``TALLRec (Book)'' on two testing sets when the number of rec-tuning samples exceeds 64. 
This finding indicates that TALLRec can seamlessly integrate data from different domains to enhance its generalization performance. In future work, it is promising to pre-train TALLRec with large-scale recommendation data from heterogeneous domains.

    

\par

\section{Related Work}
\textbf{$\bullet$ LMs for Recommendation.}
There have been several attempts to integrate language models (LMs) with recommendation systems.
Despite the incorporation of LMs~\cite{lm4rec0, lm4rec1}, some attempts persist in utilizing traditional user/item IDs to represent users/items.
Thereby,  they disregard the semantic understanding capabilities of LMs, such as reviews, which other work has incorporated the language information as part of the users/items embedding~\cite{lm4rec5}.
In addition, other methods either utilize an undisclosed model that already possesses preliminary recommendation capabilities~\cite{lm4rec2}. or employ small models to train on large-scale downstream task data~\cite{lm4rec4}.
Moreover, the aforementioned models are also limited to small models, while this paper is orthogonal about how to adapt large language models to recommendation tasks.
In recommendation systems, there is currently little research on applying LLMs in recommendation scenarios.
Those works utilize the interaction ability of GPT3.5 series models and apply In-context Learning~\cite{gao2023chat, wang2023zero}.
In detail, Chat-Rec~\cite{gao2023chat}  endeavors to harness the interaction capabilities of ChatGPT and link the ChatGPT with traditional recommendation models (e.g. MF~\cite{koren2009matrix}, LightGCN~\cite{he2020lightgcn}) to formulate a conversational recommendation system.
NIR~\cite{wang2023zero} shares a similar concept with Chat-Rec, which employs conventional recommendation models to generate candidate items subjected to a three-stage multi-step prompting process for re-ranking.

\vspace{5pt}
\noindent\textbf{$\bullet$ Sequential Recommendation.}
Our setup is close to the sequential recommendation, which aims to infer the user's next interaction based on users' historical interaction sequences~\cite{seq_survey_1, seq_survey_2}. 
In the early time, the Markov chain plays an important role in sequential recommendation~\cite{seq_markov_1, seq_markov_2, seq_markov_3, seq_markov_4}. 
Recently, deep learning-based methods have become mainstream. 
Extensive work using different kinds of neural network structures, like RNN~\cite{GRU4Rec,seq_rec_1,seq_rec_2}, CNN~\cite{Caser, CNN_seq_1, CNN_seq_2}, and attention~\cite{SASRec, attention_seq_1, attention_seq_2}, to model the user interaction sequences. However, limited by only using IDs to represent users and items, such work cannot fastly adapt and generalize to new scenarios. Thus, some works focus on the generalization ability of sequential recommendation models by pre-training~\cite{pretrain_1, pretrain_2}, data augmentation~\cite{contrast_1, contrast_2, contrast_3, contrast_4}, debiasing~\cite{debias_1, debias_2, debias_3, debias_4}, and robust optimization~\cite{group_dro_1, DROS}. 
However, they ignore the strong generalization ability of existing LLMs, leading to inadequate exploration. 

\section{Conclusion}
With the advancement of LLMs, people are gradually recognizing their potential in recommendation systems~\cite{IDvsmodality, upper_limits_of_llmrec, zhang2023chatgpt}.
In this work, we explored the feasibility of using LLMs for the recommendation. 
Our initial findings reveal that even the existing best LLM models do not perform well in recommendation tasks. 
To address this issue, we proposed a TALLRec framework that can efficiently align LLM with recommendation tasks through two tuning stages: alpaca tuning and rec-tuning. 
Our experimental results demonstrate that the LLMs trained using our TALLRec framework outperform traditional models and exhibit strong cross-domain generalization abilities. 
Moving forward, we plan to explore more efficient methods to activate the recommendation ability of larger models and tune LLMs to handle multiple recommendation tasks simultaneously.

\begin{acks}
This work is supported by the National Natural Science Foundation of China (62272437), and the CCCD Key Lab of Ministry of Culture and Tourism.
\end{acks}

\bibliographystyle{ACM-Reference-Format}
\bibliography{00_ref}










\end{document}